\documentstyle[11pt,epsfig]{article}
\def\beq{\begin{equation}}
\def\eeq{\end{equation}}
\def\beqa{\begin{eqnarray}}
\def\eeqa{\end{eqnarray}}
\def\gap{ {\ \raisebox{-.7ex}{\rlap{$\sim$}} \raisebox{.4ex}{$>$}\ }}
\def\lap{ {\ \raisebox{-.6ex}{\rlap{$\sim$}} \raisebox{.4ex}{$<$}\ }}
\begin{document}
\begin{flushright}
{\large DESY 98-164\\[1mm] 
October, 1998 } 
\end{flushright}
\vskip18mm
\begin{center}
{\LARGE Large mixing, family structure,
and \\[4mm] dominant block in the neutrino mass matrix}
\\[15mm]
{\large
Francesco Vissani\\[5mm]
{\it Deutsches Elektronen-Synchroton, DESY}\\[1mm]
{\it Notkestra\ss{}e 85, D-22603, Hamburg, Germany}}
\end{center}
\vskip10mm
\begin{abstract}
A possible connection between 
the flavour structure 
of the charged fermions and
the large $\nu_\mu-\nu_\tau$ mixing
motivates an ansatz for the
neutrino mass matrix with a dominant block.
We distinguish between a general 
form and the specific forms of the ansatz,
and concentrate on the cases 
of phenomenological interest.
The general form 
can incorporate an observable amount of
CP violation in the leptonic sector.
Only specific forms can incorporate the 
Mikheyev-Smirnov-Wolfenstein 
solutions for solar neutrinos, 
with small or large mixing angles.
Other specific variants 
explain the Los Alamos neutrino 
anomaly, or provide a two-neutrino 
hot dark matter component. 
\end{abstract}
\thispagestyle{empty}
\setcounter{page}{0}
\newpage
\section{General form of the ansatz, CP-violation}
Recently, a simple ansatz for neutrino 
masses was proposed \cite{sy,abe}:
\begin{equation}
{\cal L}_{\nu\ mass} \sim
-m_\nu\ 
(\nu_\tau,\nu_\mu,\nu_e)\ 
\left(
\begin{array}{ccc}
1 & 1 & \epsilon \\
1 & 1 & \epsilon \\
\epsilon & \epsilon & \epsilon^2 
\end{array}
\right)
\left(
\begin{array}{c}
\nu_\tau\\
\nu_\mu\\
\nu_e
\end{array}
\right)
 + {h.c.};
\label{1}
\end{equation}
$\nu_{\tau,\mu,e}$ are the neutrinos of 
a given flavour (bispinorial fields),
$\epsilon$ is a small parameter discussed below.
The neutrino mass scale $m_\nu$ can be expressed as:
$$
m_\nu=\frac{v^2}{M_{heavy}},\ \ \ \ \ \ \ \ \ v=174 \mbox{ GeV} ,
$$
where $M_{heavy}$ is a heavy mass scale, 
suggestive of a seesaw explanation 
of the smallness of the neutrino 
masses \cite{seesaw}. 
The elements of the mass 
matrix are specified up to 
(generally complex) coefficients of order of unity;
whenever these coefficients affect 
a relation, we will use the symbols 
$\sim,$ $\gap$ or $\lap$.
The ansatz is characterized by a ``dominant block''
in the $\nu_\mu,\nu_\tau$ subspace, 
in the sense that the elements of this block  
are much larger than the others, 
as accounted by the parameter $\epsilon$, 
and are comparable among them.

The underlying idea of 
this structure for the mass matrix
is to identify $\nu_\mu$ and $\nu_\tau,$ 
but to distinguish them from $\nu_e,$ 
in a model where 
known characteristics of the 
charged fermions are also reproduced. 
It was suggested \cite{sy} that in
SU(5) context, two matter fields 
$\bar 5=(L,D^c)$ may have the same flavour 
properties (nonparallel family structure, 
in the following ``NFS model'').
A  single parameter $\epsilon$ 
describes the flavour structure 
of charged fermions:
\begin{equation}
\epsilon=1/20\simeq m_\mu/m_\tau .\ \ \ 
\ \ \ \ (\mbox{NFS value}) .
\label{NFS}
\end{equation}
The same goal was achieved 
by an anomalous U(1) family symmetry 
under which $\nu_\mu$ and 
$\nu_\tau$ have the same charge
\cite{abe} (``ABE model''), 
which suggested instead the identification 
with a power of the Cabibbo angle:
\begin{equation}
\epsilon = 8\cdot 10^{-3}\simeq (\sin\theta_C)^3 \ \ \ \ 
\ \ \ (\mbox{ABE value}).
\label{ABE}
\end{equation}
The identification of the flavour properties of 
the muon and tau neutrinos should not 
be taken naively.
In fact, the ansatz is expected to be 
valid in a basis 
$\tilde \nu_e,\tilde \nu_\mu$ and $\tilde \nu_\tau$,
where the SU(2) partners have 
comparable couplings for
$\tilde \tau \tilde \tau^c$ and for  
$\tilde \mu \tilde \tau^c.$ 
Hence, a large angle rotation on the
``left'' charged leptons is necessary 
to reach the flavour basis.
Only a theory of the 
coefficients of order unity
can dissipate all the suspects 
that this rotation could spoil 
(or seriously affect) the 
ansatz; we are assuming that this does 
not happen, or, more precisely, that 
the ansatz is valid in the flavour basis
(cfr.\ with \cite{fa}.).

The simplicity  of the neutrino 
mass matrix (\ref{1}) 
is remarkable and 
perhaps of profound meaning.
We will discuss this ansatz 
and its implications
in the two models defined 
above, and in more 
generic cases (that is, treating 
$\epsilon$ and $m_\nu$ as parameters).
Our aim in particular is: 
(a) to disentangle the generic 
and specific features 
of the ansatz  (\ref{1}), 
and (b) to identify its predictions.

The mass matrix in 
eq.\ (\ref{1}) has only one massive 
eigenstate, mostly a combination of the  muon 
and tau neutrino  flavour eigenstates, whereas 
the other two states are massless.
In general, this is modified when the factors 
of order unity are taken into account.
There are three typical features:
\begin{itemize}
\item[$i)$] There are two massive 
neutrinos, $\nu_2$ and $\nu_3,$
with masses of order $m_\nu,$
and with a comparable mass splitting; 
one lighter state has
of mass order $m_\nu\cdot \epsilon^2.$
The natural assumptions for the eigenvalues of the 
``dominant block'' are therefore the
{\em no-singularity} hypothesis and the 
{\em no-degeneracy} hypothesis, the reference
scale  for ``small'' mass 
being $m_\nu\cdot \epsilon.$
\item[$ii)$] The muon and tau neutrinos are mostly
combinations of the two heavy states, with a
large mixing angle $\theta\simeq \pi/4.$
\item[$iii)$] The electron 
neutrino is almost coincident 
with the lightest state $\nu_1$, 
and has mixing angles of order 
$\theta\sim \epsilon$ with 
the heavier states 
(in another notation, $|U_{e3}|\sim |U_{e2}|
\sim \epsilon$):
\begin{equation}
\sin^2(2 \theta) \sim (2\epsilon)^2 \simeq 10^{-2}\mbox{ or }
2.5\cdot 10^{-4}, 
\mbox{ for NFS or ABE model}
\label{2}
\end{equation}
\end{itemize}

Feature $i)$ suggests that the
splitting between the two heavier 
neutrinos is $\sim m_\nu^2$
unless the {\em no-degeneracy} 
hypothesis is not fulfilled.
Therefore, assuming 
that $M_{heavy}\sim 7\cdot 10^{14}$ GeV,
it is possible to incorporate the mass splitting 
necessary for the explanation of atmospheric
neutrinos $\Delta m^2_{atm}\simeq 
2\cdot 10^{-3}$ eV$^2.$
Unfortunately, it is not possible to be more specific 
and {\em predict} whether 
$\Delta m^2_{atm}>2\cdot 10^{-3}$ eV$^2$ 
or the contrary is true
(which is of crucial importance for the
success of the K2K 
experiment \cite{K2K,K2Kbis}). 
The large mixing angle instead 
comes automatically due to $ii),$ and 
this is the most attractive 
feature of the ansatz. In fact, 
this permits to explain the atmospheric 
neutrino anomaly \cite{aa} in terms
of neutrino oscillations \cite{osc}.
The mixing (\ref{2}) is quite close to the 
small mixing angle necessary for the adiabatic 
Mikheyev-Smirnov-Wolfenstein (MSW) solution \cite{msw}
of the solar neutrino problem, 
or to the angle required to explain 
the LSND anomaly \cite{lsnd} (with NFS value). 
But, generically, also the 
mass splittings 
with the lighter neutrino
are also $\sim m_\nu^2,$ for 
the {\em no-singularity} hypothesis: 
therefore, accounting
for oscillations of atmospheric neutrinos,
the solar or LSND neutrino 
anomalies remain unexplained.
We will discuss in the following Sections 
the specific cases in which this 
conclusion does not hold.

However, one could take a radically 
theoretical point of view, and suggest 
(based on the ansatz, 
and on the mass scale chosen) 
that only the atmospheric neutrinos
{\em should} be explained  
by neutrino oscillations.
This option would probably 
become more attractive 
if the results of the 
chlorine experiments
could not be reproduced, 
the SNO experiment \cite{sno} 
failed to detect an excess of 
neutral current induced events\footnote{An 
alternative interpretation would be in 
terms of existence of
sterile neutrinos \cite{Juha}.}, 
the theoretical extimations 
for solar neutrinos were much 
more uncertain than expected,  
and the LSND anomaly was not 
confirmed by other experiments.
The appealing feature of having two 
large mass splittings 
is the possibility 
to search for CP-violation 
in the leptonic sector 
in terrestrial experiments
\cite{cplept1,cplept2},
for instance by comparing 
the probabilities of conversion:
$$
P(\nu_\mu\to\nu_e)-P(\overline{\nu}_\mu\to\overline{\nu}_e)
=4 J_{lept}\ \sin \varphi_{21} \sin \varphi_{31} \sin \varphi_{32} .
$$
The phases of oscillations in vacuum 
$\varphi_{ij}=\Delta m^2_{ij} L/(4 E_\nu)$
are comparable among them, 
and of interest for  
terrestrial experiments since: 
$\Delta m^2_{ij}\sim \Delta m^2_{atm}.$
The leptonic analogue of the Jarlskog invariant:
$$J_{lept}=\mbox{Im}(U_{\mu1}^* U_{e1} U_{\mu2} U_{e2}^*)$$
is suppressed due to the presence of 
two small mixings (eq.\ (\ref{2})):
$J_{lept}= {\cal O}(\epsilon^2)$ 
(or smaller, depending on the phases). 
Observable effects in forthcoming experiments 
would imply:
$$
\epsilon \gap 0.1,
$$
that is not far from the value
in the NFS model (\ref{NFS}), but 
much larger than the value (\ref{ABE}).

Another feature of the models (\ref{NFS}) and (\ref{ABE}), is 
that the electron neutrinos are not expected 
to undergo any significant oscillation in the atmosphere.
Hence, the search for a subdominant mixing by 
using the lowest energy
neutrino-induced events \cite{ue3,par,parr,v}
will also test the mass matrix (\ref{1}).
The present bound $|U_{e3}|^2<0.15$ at 90 \% CL, obtained
in a model with a single neutrino $\nu_3$ splitted in mass
\cite{ue3}, suggests the limit:
$$
\epsilon^2 \lap 0.1.
$$
This is far also from (\ref{NFS}).
(Notice that we {\em estimated} the sensitivity 
of the atmospheric neutrino studies to 
small mixings; but, in order to assess precise bounds,
it is necessary to go beyond the usually \cite{ue3,par,parr,v}
kept hypothesis 
of degenerate $\nu_1$ and $\nu_2.$) 

We offer a theoretical guess for a large 
value of $\epsilon,$ compatible with both 
previous bounds, and suggestive of a link between
the neutrino mass matrix and the flavour structure:
\begin{equation}
\epsilon = 0.2 \simeq
\sin\theta_C  \ \ \ \ \ \ \ \ (\mbox{``large'' value}) .
\label{verylarge}
\end{equation}
A component of oscillation $|U_{e3}|$ 
of this size could be detectable, 
studying the dependence on the energy cuts of the 
observed event rates at the Super-Kamiokande 
(SK) \cite{parr}.

\section{Specific forms of the ansatz: Solar neutrinos}

Alternatively, we are led to the conclusion
that only quite specific 
versions of the ansatz (\ref{1})
for the neutrino masses, 
where feature $i)$ is not valid, can 
reconcile the indications of 
atmospheric neutrino oscillations
with other neutrino anomalies.
We are interested in
the specific subset of the
mass matrices of type (\ref{1})
that predict the existence of only 
one neutrino with mass of order $m_\nu.$ 
We relax the 
hypothesis of {\em non-singularity} 
in the ``dominant block'',
introducing a new small parameter $\delta,$ such that
$${\cal M}_{\tau\tau}/{\cal M}_{\mu\tau}= 
{\cal M}_{\mu\tau}/{\cal M}_{\mu\mu} + {\cal O}(\delta);$$
now, an eigenvalue of the 
dominant block is order 
$m_\nu\cdot \delta.$ 

On the theoretical side, 
we remark that a relatively high degree 
of tuning is a valuable information on 
the flavour structure, that goes beyond 
what is suggested by the NFS or ABE models,
and in general by the ansatz considered.
Keeping in mind the seesaw \cite{seesaw} 
formula for the light neutrino masses,
${\cal M}= - v^2 Y_\nu^t M_R^{-1} Y_\nu,$ 
such a  tuning may 
suggest us that the determinant 
of (a block of) the neutrino Yukawa couplings $Y_\nu$
is (close to) zero, cfr.\ \cite{fa},
or that at leading order  only one eigenvalue of the  
right-handed neutrino mass $M_R$ is sufficiently 
light to contribute to ${\cal M},$ cfr.\ \cite{bhs}.
Eq.\ (10) in \cite{fa} (second paper)
or ``texture $I\!I$'' in \cite{bhs}, 
are equivalent to the mass matrix we
are considering, in the limiting case
$\epsilon=\delta=0$.
Indeed, in \cite{fa}
the connection between neutrino 
masses and the flavour problem
is also considered. 
Mechanisms to adequately 
control the flavour structure
of the neutrino mass matrix 
have also been proposed in 
\cite{gns}.

After changing basis in the 
dominant block, from 
$(\nu_\tau,\nu_\mu)$
to the ``heavy'' and ``light'' states
$(\nu_h,\nu_l),$
the neutrino mass matrix becomes:
\begin{equation}
{\cal M}'\sim 
m_\nu\ 
\left(
\begin{array}{ccc}
1 & 0 & \epsilon \\
0 & \delta & \epsilon \\
\epsilon & \epsilon & \epsilon^2 
\end{array}
\right) ,
\label{3}
\end{equation} 
where we assumed that the rotation is not 
a source of suppression for the 
${\cal M}'_{el}$ and ${\cal M}'_{eh}$ entries
({\em no-special-zeroes} hypothesis).
We notice that the heaviest neutrino state, $\nu_h,$
can be integrated away, leaving us 
a two-flavour task; the only effect is that the
${\cal M}'_{ee}$ entry receives another  
contribution, but still
of order $\epsilon^2.$ 
Like for eq.\ (\ref{2}), the mixing 
of the heaviest neutrino state with 
the electron neutrino is of 
order $\epsilon.$ 
The discussion related to
eq.\ (\ref{verylarge}) 
is still valid.

Oscillations of atmospheric  neutrino 
set the mass scale in eq.\ (\ref{3}) to
$m_\nu^2\sim \Delta m^2_{atm}.$
Now we study the possibility 
of incorporating also solar neutrino oscillations,
taking advantage of the parameter $\delta.$
For this sake, we 
first rewrite the effective mixing 
matrix between the light states, 
writing explicitly 
the coefficients of order unity, $p,q$ and $r$:
\begin{equation}
{\cal M}_{2\times 2}=
\sqrt{\Delta m^2_{atm}}\times
\left(
\begin{array}{cc}
p\ \delta, & r\ \epsilon \\
r\ \epsilon, & q\ \epsilon^2
\end{array}
\right).
\label{mmm}
\end{equation} 
($q$ is inclusive of the heavier 
neutrino ``seesaw'' contribution).
The relation between the elements of the 
neutrino mass matrix
and the parameters of 
oscillation in vacuum is:
\begin{equation}
\left\{
\begin{array}{l}
\delta m^2\ \sin 2\theta= 2\ \Delta m^2_{atm}\
|r| \ \epsilon\ |p^*\ \delta\ +q\ \epsilon^2|\\[2ex]
\delta m^2\ \cos 2\theta= \Delta m^2_{atm}\ 
(|p|^2\ \delta^2  -|q|^2\ \epsilon^4)  
\end{array}
\right. ;
\label{pars}
\end{equation}
$\delta m^2\ge 0$ is non-negative,
and $\theta\in [0,\pi/2].$ 
The propagation in a medium with electrons 
density $\rho_e$ effectively modifies 
the second term: 
$\delta m^2_{\mbox{\it eff}}\ \cos 2\theta_{\mbox{\it eff}}= 
\delta m^2\ \cos 2\theta - 2 \sqrt{2}\ E_\nu\ G_F\ \rho_e.$  
Thus, to have maximal $\theta_{\mbox{\it eff}}$ due to 
the interplay of vacuum and matter terms 
\cite{msw} we need $\delta \gap \epsilon^2.$ 
Hence, at $\delta=0,$ the 
MSW mechanism does not work---or in other terms, 
the case when the singularity of the 
dominant block is exact (or very pronounced) 
is a special one.

The allowed values of 
the oscillation parameters
are delimited by the two curves:
\begin{equation}
\delta m^2 =\Delta m^2_{atm}\ \epsilon^2
\times \frac{4\ |r|^2}{\sin^2 2\theta}\times
\left[ \cos 2\theta \pm 2 \epsilon\ \frac{|q|}{|r|} \sin2\theta \right] .
\label{reg}
\end{equation}
These regions are 
represented in figure 1
for two values of $\epsilon.$
Allowing for an excursion of a factor 
of 4 in $\Delta m^2_{atm},$ 
and a factor of 2 in the moduli of the 
coefficients $q$ and $r$ in previous equation,
the region extends by more than two orders 
of magnitude in $\delta m^2$
for any assigned value of $\theta.$
The regions close to $\theta=\pi/4$ correspond
to  $\delta$ comparable to $\epsilon^2$
(or smaller). 
The presence of a large mixing is 
evident from eq.\ (\ref{mmm}), since for 
such a small $\delta$ the two neutrino states form a 
quasi-Dirac neutrino of mass $m_\nu\cdot \epsilon$
(degeneracy being broken at 
next order in $\epsilon$).
In this region, the range 
of values taken by $\theta$ 
for fixed $\delta m^2$ is due to the size
of $\epsilon,$ from the bracketed term 
in eq.\ (\ref{reg}); 
the shrinking from  NFS to the ABE 
value of $\epsilon$ 
results clearly in the figure.
The transition from the 
$\theta\simeq\pi/4$ region, 
to the ``typical'' small 
mixing angles, eq.\ (\ref{2}), 
takes place for increasing $\delta$'s. 

The large mixing angle solutions 
of the solar neutrino problems 
are easy to obtain.

But it is possible to do even better, and
reproduce the small mixing angle 
MSW solution (SMA) 
at best fit values \cite{bsk} if: 
\begin{equation}
\epsilon \sim  \sin2\theta\ \left(
\frac{\Delta m^2_\odot}{ 4\ \Delta m^2_{atm}}
\right)^{1/2}
= 2\cdot  10^{-3} 
\label{best}
\end{equation}
(the preliminary SK
data on the shape of the 
electron spectrum tend to disfavour the
large angle solutions \cite{suzuki,bsk}).
The value in (\ref{best}) is close to the value 
obtained in the ABE model,
but much smaller than in the NFS 
model (see figure 1).
It is very well consistent with:
\begin{equation}
\epsilon=(\sin\theta_C)^4\ \mbox{or}\ (m_\mu/m_\tau)^2 \ \ \ \ \ \
\ \ \ (\mbox{``small'' value}) .
\label{small}
\end{equation}
Correspondingly to eq.\ (\ref{best}), we have 
$$\delta\sim 5\ \cdot 10^{-2}\simeq (\sin\theta_C)^2.$$
This result, $\delta \sim \sqrt\epsilon,$
tells us that to obtain the SMA solution
the breaking of the hypothesis of {\em no-singularity}
must be weaker than $\epsilon$ 
(more in general, this last 
condition implies small mixing angles, 
as it is apparent from eq.\ (\ref{mmm})).

The implementation of 
the small angle solution 
is not possible in the 
NFS model (\ref{NFS}) without 
special arrangement in the 
parameters of order unity: 
a suppression of $r$ by 
a power of $\epsilon$ ($\sim 1/20;$
see  eqs.\ (\ref{pars})) 
is necessary to hit SMA solution.
It is possible in principle that 
a specific implementation of the 
NFS model \cite{sy} produces a violation 
of the {\em no-special-zeroes} hypothesis, 
but we don't see the reason for that
in the present context.

It is interesting to notice that, since maximal
mixing ($\sin^2 2\theta\simeq 1$) 
is permitted, completely averaged 
neutrino oscillations with 
survival probability $P(\nu_e\to\nu_e)\simeq 1/2$
\cite{ave} are possible.
Many experimental informations on solar neutrinos
can be accounted for in this assumption, 
with the noticeable exception of the counting rate 
of the chlorine experiment \cite{davis} and of
the spectral shape at SK
(a detailed study is in \cite{conforto}).
For $\delta m^2$ 
smaller than those represented 
in the figure, it is possible to reproduce 
also the vacuum oscillation solution,
sometimes called just-so \cite{p}.

Let us now try to discuss the 
likelihood of the possible 
solutions we found.
This of course requires to make 
a priori assumptions, 
that are however necessary 
in order to make the model predictive.
In this respect, we remark that if we treat both
$\epsilon$ and $\delta$
as free parameters, {\em any} solar 
neutrino solution can be fitted;
so we take as references the 
NFS (\ref{NFS}) or ABE (\ref{ABE})
values of $\epsilon$ in the following discussion.
A criterion for absence of 
fine-tuning is that $\delta$ 
should be {\em at most} 
of order $\epsilon,$ that 
leads to a lower limit on $\delta m^2$: 
\begin{equation}
\delta m^2 \gap \Delta m^2_{atm}\ \epsilon^2.  
\end{equation}
This condition already gives us  a lot of freedom: 
For the NFS model, 
the large angle mixing (LMA) is possible;
for the ABE model the SMA solution is possible.
More specifically, if the origin of the 
smallness of $\delta$ and of $\epsilon$
is the same, we may expect 
$\delta\sim \epsilon;$
and using eq.\ (\ref{pars}),
we conclude that large angles are 
to be expected in this assumption.
The case $\delta\sim \epsilon$
is interesting, since we 
can reproduce the LMA 
solution with a value 
of $\epsilon$ close to the one of 
the NFS model (\ref{NFS}).
Now, let us consider values 
$\delta \lap \epsilon.$ 
Even in this case, we 
meet a second (but of course weaker) 
fine tuning criterion.
As visible in figure 1, it is possible 
to obtain $\delta m^2$ as small as desired,
in the region where the two 
curves become vertical. 
This requires, however, a special 
arrangement between the phases 
and moduli of $p$ and $q,$
and $\delta\sim \epsilon^2,$ 
as one verifies from  eqs.\ (\ref{pars}).
The border of the region 
of parameter space in which 
it is not necessary to admit 
this delicate fine tuning 
can be estimated 
assuming $\delta=0.$ 
Eqs.\ (\ref{pars}) suggest then:
\begin{equation}
\delta m^2 \gap 2\ \Delta m^2_{atm}\ \epsilon^3 
\label{ft}
\end{equation}
Now, also the solution denoted as LOW in
\cite{bsk} can be incorporated 
by both NFS and ABE models (figure 1);
similarly, it is possible to  find agreement with 
the low $\delta m^2$ (and large angle)
solutions that are suggested by the 
analysis of the SK data alone \cite{suzuki}.

Averaged solutions are ``more natural'' for large
values of $\epsilon$ (like those in eq.\ (\ref{verylarge})). 
Quite small $\epsilon$'s, or more fine-tuning, are required 
to get vacuum oscillation solutions.

Summarizing, the ``specific'' form of the  
ansatz--eq.\ (\ref{3})--is able to incorporate 
MSW solutions of the solar neutrino problem,
with large (NFS model) or small mixing angle 
(ABE model), or even other possibilities.
We conclude that, 
in order to state actual 
predictions, precise values 
should be given not only for $\epsilon,$
but also for the overall 
mass scale $m_\nu,$ for $\delta$ and for the 
coefficients of order unity.

\section{Specific forms of the ansatz:
LSND anomaly, dark matter}

Let us consider now a 
second possibility, in 
which we simply neglect 
the solar neutrino problem.
In the case in which 
the mass scale $m_\nu$ is 
large, we can still take advantage 
of the presence of large 
mixing angles in the ansatz (\ref{2})
and explain the
atmospheric  neutrino problem in terms of 
oscillations, but now a fine tuning 
must be operated on the parameter $\eta$:
$$
\eta=\frac{\Delta m^2_{atm}}{m_\nu^2} .
$$
As in Section 2, we 
assume that this arises as a sort 
of fine structure of the 
dominant block.
In the terminology of 
Section 1, we are specifying the 
general ansatz (\ref{1}) relaxing 
the {\em no-degeneracy} hypothesis.

In the first variant, we can use 
the small mixing angles (\ref{2}) 
of the electron neutrino with the
heavy states to explain 
the LSND anomaly \cite{lsnd}. 
This can be done in the NFS model,
if $m_\nu\sim 0.7$ eV
($M_{heavy}\sim 4\cdot 10^{13}$ GeV),
that implies for the fine tuning parameter 
$\eta=4\cdot 10^{-3}.$
It is not possible, instead, 
to account for LSND anomaly 
in the ABE model, since 
the angle is too small.

In the second variant,
we consider a larger mass 
scale, $m_\nu\sim 2.5$ eV,
related to two-neutrino hot 
dark matter component (2$\nu$HDM), 
$\Omega=1$ cosmological model \cite{primack}. 
This assumption is not consistent 
with the bounds from reactor 
experiments \cite{karmen} in the 
model NFS ($\sin^22\theta$ should
be reduced by one order of magnitude),
whereas the ABE value for $\epsilon$
is fine, again because of
the small mixing angle.
The parameter $\eta$ and 
the heavy scale $M_{heavy}$ 
are smaller than in the previous case.

\section{Conclusions and perspectives}
The principle of 
connecting neutrino masses to 
family structure is very attractive.
A related ansatz for 
massive neutrinos was 
discussed, both in 
its generic form (\ref{1}) 
(with parameters $m_\nu$ and $\epsilon$)
and in its specific form (\ref{3})
(with the additional parameters 
$\delta,$ or $\eta$).

The clearest feature of the ansatz 
considered is about atmospheric neutrinos. 
Due to the smallness of $\epsilon,$
almost pure $\nu_\mu-\nu_\tau$ 
oscillations take place.
Many relevant tests will be 
possible in the relatively short term:\\ 
1) The analysis of low energy induced 
stopping muons, and partially contained events
(presented in preliminary form by the 
MACRO and SK collaboration \cite{aa}).\\
2) The search of  ``oscillated'' 
$\nu_\tau$ flux via neutral current 
induced events \cite{nc}. 
To perform this inference from observed events, 
a detailed knowledge of the low energy 
neutrino interactions is required\footnote{New 
experimental informations 
will  be obtained already at the ``close'' 
detector of the 
K2K experiment \cite{K2K,K2Kbis}.}.
A failure of the detection 
would rule out the model.\\
3) The studies of a subdominant 
electron neutrino component in 
atmospheric neutrinos, and 
further reactor studies.
Both should give null result, {\em except
if $\epsilon$ is large, as in eq.}\ (\ref{verylarge}).\\
4) A cross-check, or in any case  
an improvement of our 
knowledge on the parameter
$\Delta m^2_{atm}$ in 
long baseline experiments.\\
5) The search of charged current induced
$\tau$'s, produced either 
in artificial beams, or by the atmospheric neutrinos
themselves (for some elements for a discussion, 
see \cite{v}).\\
The theoretically weak point of the model is that
the scale $M_{heavy}$ is fitted, not predicted.
However, it is appealing that the scale is $1-2$ 
orders of magnitude  smaller than the (minimal SU(5))
supersymmetric grand unification scale
(in the variants considered in the 
previous Section, it is even smaller). 
Since, in a predictive model for the scale,
there are typically perturbative couplings $y$ ($y<1$) 
such that $v^2/M_{heavy}\equiv y^2 v^2/M_X,$ 
the actual scale $M_X$ is expected to 
be lower than $M_{heavy}.$
Hence, there is {\em a strong suggestion for the 
existence of an intermediate mass scale,} 
possibly right handed neutrinos,
even if this is not an unescapable conclusion
(for instance, neutrino masses may be related
to a ``direct'' mass term \cite{seesaw2}).

We summarize in table 1 
other observable 
features that the ansatz
can incorporate,
using eqs.\ (\ref{NFS}) and 
(\ref{ABE}) for $\epsilon,$  
and using the additional small parameter 
$\delta$ in second to fourth cases,
and $\eta$ in the last two.
Other interesting values of
$\epsilon$ have been discussed 
in (\ref{verylarge}) and (\ref{small}).
\begin{table}[btt]\label{summary}
\begin{center}
\begin{tabular}{|c||c|c|c|c|c|c|c|} 
\hline
  & $\mbox{\footnotesize CP}\!\!\!\!/$ & $\nu_\odot$ 
{\footnotesize SMA} & $\nu_\odot$ {\footnotesize LMA} 
& $\nu_\odot$ {\footnotesize LOW}  
& $\nu_\odot$ {\footnotesize aver.} & {\footnotesize LSND} 
& 2$\nu${\footnotesize HDM}  \\  
\hline
ABE & no & yes &  no & yes & yes & no & yes \\
\hline
NFS & $\sim$yes & no & yes & yes & yes & yes & $\sim$no \\
\hline
\end{tabular}
\end{center}
\caption{Alternative facts that 
can be incorporated by the models 
NFS (\ref{NFS}) and ABE (\ref{ABE})
together with atmospheric 
neutrino oscillations.}
\end{table}
The explanation 
in terms of massive neutrinos 
of some fact, 
among 1) solar neutrino problem, 
2) LSND anomaly or 3) dark matter 
neutrinos,
will preclude the explanation 
of the other ones: 
The model can be brought to 
logical contradiction if 
two (or three) facts were proved 
to be correct. 
Among the particular implications, 
it is interesting that 
the small mixing angle MSW solution 
can not be implemented in the 
NFS model without fine-tunings.
Models with observable CP violation
(in which the three indications cannot be
accounted) are also possible.

To increase the predictivity 
of the ansatz, a detailed theory of 
the coefficients of order unity, 
and particularly of the structure of 
the dominant block is necessary.

\vskip1truecm

It is a pleasure to 
thank Riccardo Barbieri 
and Ferruccio Feruglio
for helpful correspondence,
Ahmed Ali, Wilfried Buchm\"{u}ller, Eligio Lisi
and Goran Senjanovi\'c for useful 
discussions, and Sandro Ambrosanio 
for an important remark.

\newpage

\begin{figure}[htb]
\centerline{\epsfig{file=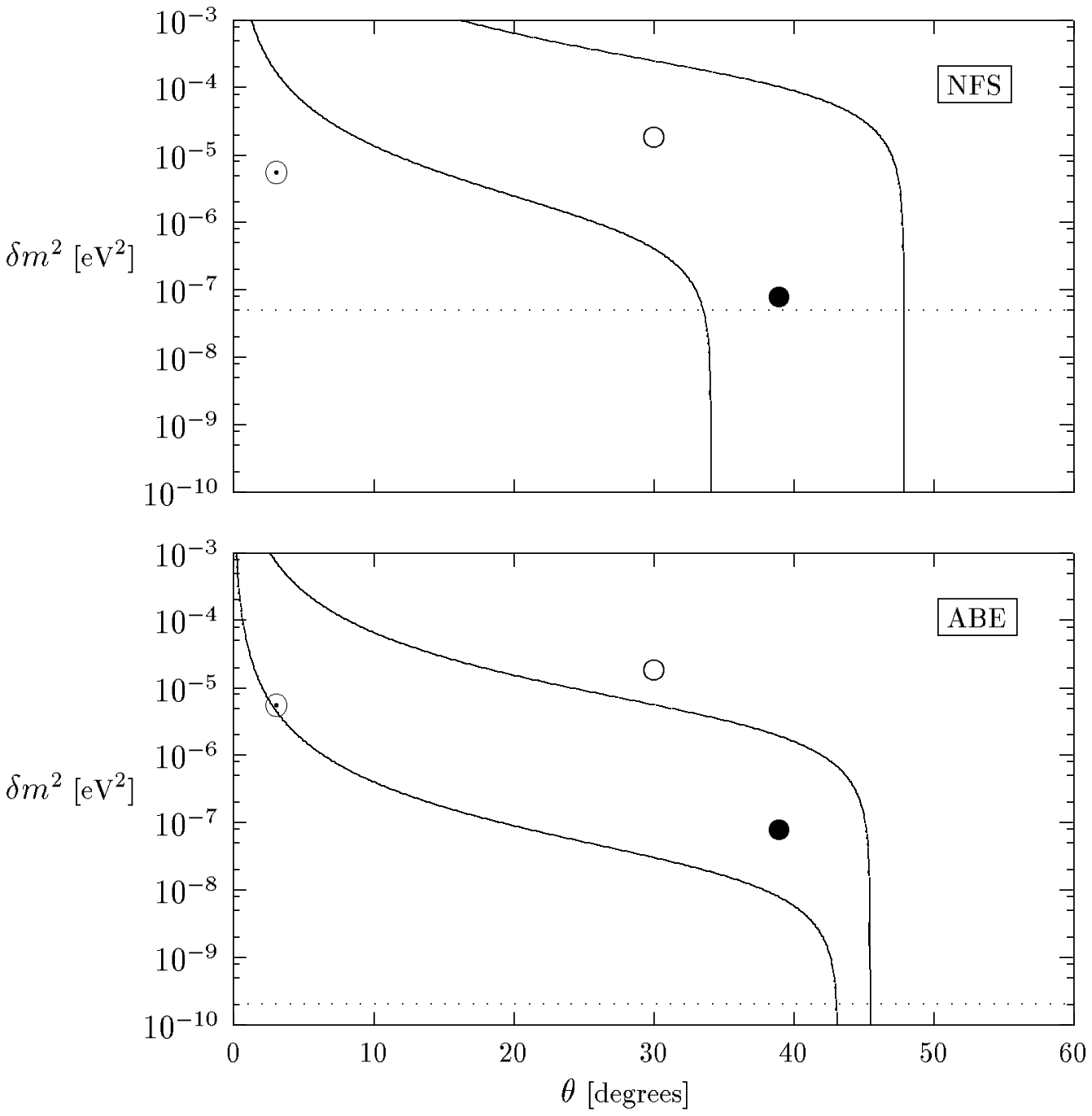,width=15cm}}
\caption{Allowed regions for 
solar neutrino parameters 
in the NFS and ABE models. 
Observed counting rates in
solar neutrino experiments 
can be explained in the vicinity of 
the indicated points \cite{bsk}: 
dotted, void and filled circles correspond
to (best fit) SMA, LMA and LOW ``solutions''.
The dotted line denotes 
the lower value of $\delta m^2$ 
compatible with the 
fine tuning condition (\ref{ft}).}
\end{figure}

\end{document}